\documentclass[11pt]{article}
\usepackage{graphicx,hyperref}
\usepackage{color,graphicx,amsmath,amssymb,
bm}

\newtheorem{prop}{Prop}[section]

\newtheorem{lemma}[prop]{Lemma}

\newtheorem{theorem}[prop]{Theorem}

\usepackage{makeidx}
\makeindex
\usepackage{ascmac}
\usepackage{pb-diagram}
\usepackage{here}
\usepackage{esint}
\usepackage{color}
\title{Gauge theory for quantum XYZ spin glasses }
\author{C. Itoi$^1$  and Y Sakamoto$^2$\\
\\
\vspace{3mm}
$^1$ Department of Physics,   
GS $\&$ CST, 
Nihon University
\\ %
$^2$Laboratory of Physics, CST, 
Nihon University
}

\begin{document}
\maketitle
%
\vspace{10pt}
\begin{abstract}{ Nishimori's gauge theory is extended to the quantum XYZ $p$-spin glass model in finite
dimensions.
This enables us to obtain useful correlation equalities, which show   
also that Duhamel correlation functions at an arbitrary temperature 
are bounded by those in the corresponding classical model on the Nishimori line.
These bounds give that the spontaneous magnetization vanishes
in any low temperature even if the model enters the $\mathbb Z_2$-symmetry broken spin glass phase.
This theory explains well-known fact from experiments and numerical calculations 
that the magnetic susceptibility does not diverge in the spin glass transition.
The new gauge theory together with the known phase diagram of the Edwards-Anderson model
can specify the spin glass region in the coupling constant space of the quantum Heisenberg XYZ spin glass 
model. }
\end{abstract}

%
\vspace{2pc}
\noindent
{\it Keywords}: Gauge theory, Phase diagram, Paramagnetic and Spin glass phases, Spin glass transition, 
Spontaneous symmetry breaking, Spontaneous magnetization, Magnetic susceptibility,
%
%
%
%

\maketitle
\section{Introduction}   
Phase transitions in spin glasses have been studied extensively.
In particular, the response of spin glasses to magnetic field has been investigated mathematically 
\cite{AW,GAL,AGL}, numerically \cite{B} and experimentally \cite{NKH,TM}. 
It is well-known that the static magnetic susceptibility does not diverge in the spin glass transition \cite{B,NKH}.
This fact is  consistent with rounding effects explained mathematically \cite{AW,GAL,AGL}.  

In the present paper, a new extension of Nishimori's gauge theory to the quantum  XYZ $p$-spin glass model is provided.
This theory enables us to discuss the finiteness of the magnetic susceptibility in the spin glass transition of
mathematically. Nishimori's gauge theory is well-known as a useful theory to study macroscopic behaviors of
mainly Ising spin glass systems \cite{N,N0}. The gauge invariance of Hamiltonian and gauge covariance of distribution functions of random coupling constants
provide several useful exact relations on the Nishimori line lying in paramagnetic and ferromagnetic phases
in the coupling constant space of the Edwards-Anderson (EA) model \cite{EA}. 
These are helpful to draw the phase diagram of the EA model.
It is well-known that fluctuation of order parameters is suppressed on the Nishimori line. 
Nishimori and Sherrington have argued the absence of replica symmetry breaking on the Nishimori line in the EA model \cite{NS}.  
Recently, this fact is confirmed in Nishimori's gauge theory \cite{IS2}
combined with Aizenman-Contucci-Ghirlanda-Guerra identities \cite{AC,GG,CG,CG2,C1,CG3}.
Okuyama and Ohzeki obtain
the Gibbs-Bogoliubov inequality for the free energy in the Sherrington-Kirkpatrick (SK) model \cite{SK} 
using its local concavity on the Nishimori line \cite{OO}.  
Quite recently, they have proven also that 
 the free energy of Ising spin glass models with the Kac potential 
in the non-additive limit is exactly the same as that of the SK model on the Nishimori line
 in the thermodynamic limit \cite{OO2}.

On the other hand, studies on quantum spin glass systems \cite{GAL,AGL,CGP,MON,Cr,CL,I2,I,W,IISS,IS} are
much fewer than those on classical Ising spin glass systems, since 
the non-commutativity of operators yields considerable complication for evaluations.
Morita, Ozeki and Nishimori have shown that Nishimori's gauge theory is useful also for quantum spin glasses \cite{MON}.
Correlation functions in the transverse field  EA model and the XY gauge glass at an arbitrary temperature
are bounded by those in the corresponding classical Ising spin glass  model on the Nishimori line \cite{MON}. 
These bounds yield that the absence of ferromagnetic long-range order on the Nishimori line in the classical Ising spin glass model implies 
the absence of ferromagnetic long-range order in the corresponding quantum systems. 
Although the Nishimori line lies out of the
spin glass phase, the gauge theory provides bounds on physical quantities
in spin glass phase in classical and quantum spin glass models. 
Recently, absence of spontaneous magnetization 
and boundedness of the magnetic susceptibility in the spin glass phase have been shown
in the  transverse field EA model \cite{IS}.  
     
In the present paper, several useful 
equalities among the correlation functions in quantum XYZ $p$-spin glass are
obtained by extended Nishimori's gauge theory. 
The sample expectation of a correlation function in the quantum spin glass model at arbitrary temperatures is equal to
the correlation function times another correlation function in the corresponding classical Ising spin glass model on the Nishimori line. 
These correlation equalities enable us to prove well-known properties of the spontaneous $\mathbb Z_2$-symmetry breaking
in quantum spin glass systems. It is proven that there is no spontaneous magnetization in a certain subspace of the coupling constant space 
as in the classical spin glass models.
The spontaneous magnetization in the quantum spin glass model for any temperature is bounded by those on the Nishimori line in the corresponding classical model.
 In addition, two acceptable assumptions and Nishimori's gauge theory enable us to discuss
the finiteness of ferromagnetic susceptibility in spin glass phase transition. 
A bound on the ferromagnetic susceptibility is given by a correlation function 
in the corresponding classical model on the Nishimori line. 
Under these assumptions, it is proven that 
the ferromagnetic susceptibility under zero external field does not diverge
in the paramagnetic and spin glass phases, where the spontaneous magnetization is proven to be zero by the obtained gauge theory.   
These  results are well-known general properties of spin glasses, which are valid also in the classical Ising spin glasses. 
Finally, we study the quantum Heisenberg XYZ spin glass model in the new gauge theory in combination with
the known phase diagram of the EA model. This allows us to specify a region in the coupling constant space
of the XYZ model, where the gauge theory proves that there is no spontaneous magnetization
at any temperature. The spin glass state should appear in this region at sufficiently low temperature.

 \section{Definitions of the quantum XYZ spin glass model}
 For a positive integer $L$,  let  $\Lambda_L:= [0,L-1]^d \cap {\mathbb Z}^d
$ be a $d$ dimensional cubic lattice whose volume is $|\Lambda_L|=L^d$. 
 A sequence of spin operators  $\bm \sigma:=(\sigma^{w}_i)_{w=x,y,z, i \in \Lambda_L}$
on a Hilbert space ${\cal H} :=\bigotimes_{i \in \Lambda_L} {\cal H}_i$ is
defined by a tensor product of the Pauli matrix $\sigma^w$ acting on ${\cal H}_i \simeq {\mathbb C}^{2}$ and unities.
These operators are self-adjoint and satisfy the commutation relations
$$
 [\sigma_k^y,\sigma_j^z]=2i \delta_{k,j} \sigma_j^x ,\ \ \  \ \ 
[\sigma_k^z,\sigma_j^x]=2i \delta_{k,j} \sigma_j^y ,\ \  \ \ \ [\sigma_k^x,\sigma_j^y]=2i \delta_{k,j} \sigma_j^z ,  
$$
and each spin operator satisfies
$$
(\sigma_j^w)^2 = {\bf 1}.
$$
Denote a product of  spins $$
\sigma_X^w= \prod_{i \in X} \sigma_i^w,
$$ 
for a finite sub-lattice $X \subset \Lambda_L$.
Let $p$ be  a positive integer. 
To define a short-range $p$-spin Hamiltonian, define a collection ${\cal A}_p$ of
 interaction ranges $A_p \subset \Lambda_L$,  such that
$(0, \cdots, 0) \in  A_p$ and $|A_p| =p$.  
  Define a  collection ${\cal B}_p$ of interaction ranges  by
\begin{equation}
{\cal B}_p:= \{X \subset \Lambda_L | X= i+ A_p, i\in \Lambda_L, A_p\in {\cal A}_p \}.
\end{equation}
Let ${\cal P}$ be a finite set of  positive integers. ${\cal P}$ defines 
a Hamiltonian of short-ranged mixed $p$-spin interactions by
\begin{equation}
H(\bm \sigma, \bm J) :=-\sum_{p \in {\cal P}} \sum_{X \in {\cal B}_p} \sum_{w=x,y,z} J^{w}_{X,p}
\sigma_{ X}^w,
\label{Hamilq}
\end{equation}
where,  a sequence 
$\bm J:=(J_{X,p}^w)_{p\in {\cal P},X \in {\cal B}_p,w=x,y,z}$ consists  of  independent Gaussian 
random variables (r.v.s) with its expectation value  $\mu_p^w >0$ and its standard deviation $\Delta_p^w> 0$.
The probability density function of each $J_{X,p}^w$ is given by
\begin{equation}
P_p^w(J_{X,p}^w) := \frac{1}{\sqrt{2\pi} {\Delta_p^w}} \exp \Big[-\frac{(J_{X,p}^w-\mu^w_p)^2}{2 {\Delta_p^w}^2}\Big].
\label{PJ}
\end{equation}
${\mathbb E}$ denotes the sample expectation over 
all $J_{X,p}^w$, such that  
$$
{\mathbb E} J_{X,p}^w=\mu_p^w, \ \ \ {\mathbb E} (J_{X,p}^w -\mu_p^w)^2={\Delta_p^w}^2.
$$
\begin{equation}
U_wH(\bm \sigma, \bm J) U_w^\dag=H(\bm \sigma, \bm J),
\end{equation}
for $U_w :=\sigma_{\Lambda_L }^w$ for each $w \ (=x,y,z)$, 
if  $\Delta_{p}^u=\mu_p^u=0$ for all odd $p \in {\cal P}$ for $u\neq w$.

Define Gibbs state for the Hamiltonian.
For a positive $\beta $ and  real numbers $J_{X,p}^{w}$,  the  partition function is defined by
\begin{equation}
Z_L(\beta, \bm J) := {\rm Tr} e^{ - \beta H(\bm \sigma,\bm J)},
\end{equation}
where the trace is taken over all basis in the Hilbert space.

For a certain fixed $w \ (=x,y,z)$, regard $\bm J^w$ as a set of classical couplings 
and symbolically $\bm Q^w:=(\bm J^u, \bm J^v) $ for other components satisfying $ v\neq u \neq w \neq v$
denotes a set of quantum perturbations. Represent couplings by  
$(\bm J^u, \bm J^v, \bm J^w)=(\bm Q^w, \bm J^w)$. 
Let $f$ be an arbitrary function 
of spin operators, and  the  expectation of $f(\bm \sigma)$ in the Gibbs state is given by
\begin{equation}
\langle f(\bm \sigma) \rangle^{(\bm Q^w, \bm J^w)}_\beta
:=\frac{1}{Z_L(\beta, \bm J)}{\rm Tr} f( \bm \sigma)  e^{ - \beta H(\bm \sigma , \bm J )}.
\end{equation}
Note that the index $\bm J=(\bm Q^w, \bm J^w)$ of the Gibbs expectation includes the quantum perturbation in the Hamiltonian.
Note that the Gibbs expectation of operators $\bm \sigma^w:=(\sigma_i^w)_{i\in \Lambda_L}$ 
at $\bm Q^w= \bm 0$ is identical to the Gibbs expectation in the classical  model with random exchanges $\bm J^w$
$$
\langle f(\bm \sigma^w) \rangle^{(\bm 0, \bm J^w)}_\beta =\langle f(\tau) \rangle_\beta,
$$ 
where the classical spin configuration $\tau: \Lambda_L \to \{1,-1\}$ is a function from lattice to a set of eigenvalues of $(\sigma_i^w)_{i\in \Lambda_L}$.
Duhamel function of two functions $f(\bm \sigma), g(\bm \sigma)$ of spin operators
is defined by
\begin{equation}
(f(\bm \sigma), g(\bm \sigma))^{\bm J}_\beta:=\int_0^1dt \langle e^{t\beta  H(\bm \sigma, \bm J)}  f(\bm \sigma)e^{-t \beta  H(\bm \sigma, \bm J)} g(\bm \sigma) \rangle^{\bm J}_\beta
\end{equation}
Note that
$$
(f(\bm \sigma^w), g(\bm \sigma^w))^{(\bm 0, \bm J^w)}_\beta =\langle f(\tau) g(\tau)\rangle_\beta,
$$ 
We define the following functions of  $(\beta,  \bm \Delta , \bm  \mu) \in [0,\infty)^{1+6|{\cal P}|} $ and randomness
$\bm J=(J_{X,p}^{w})_{p \in {\cal P},X \in {\cal B}_p, w =x,y,z}$
\begin{equation}
\psi_L(\beta,\bm J) := \frac{1}{|\Lambda_L|} \log Z_L(\beta, \bm J), \\ 
\end{equation}
$-\frac{L^d}{\beta}\psi_L( \beta,\bm J)$ is called free energy in statistical physics.
Define a function $p_L:[0,\infty)^{1+6|{\cal P}|} \rightarrow {\mathbb R}$ by
\begin{eqnarray}
p_L(\beta, \bm \Delta, \bm \mu):={\mathbb E} \psi_L(\beta, \bm J).
\end{eqnarray}

The following infinite volume limit
\begin{eqnarray}
p(\beta, \bm \Delta, \bm \mu):=
\lim_{L\to\infty}p_L(\beta, \bm \Delta, \bm \mu),
\end{eqnarray}
exists for each $(\beta, \bm \Delta, \bm \mu)$ \cite{I2}.
Note that the function $\psi_L(\beta, \bm J)$, $p_L(\beta, \bm \Delta, \bm \mu)$  and $p(\beta, \bm \Delta, \bm \mu)$ 
are convex functions of each variable.

To study $\mathbb Z_2$-symmetrry and its breaking, 
define order operators in terms of $w$ component of the Pauli operators 
\begin{equation}
o^{w} := \frac{1}{|\Lambda_L|} \sum_{i \in \Lambda_L} \sigma_i^w.
\end{equation}
For arbitrary functions $f(\bm \sigma), g(\bm \sigma)$ of spin operators,
denote 
their  truncated  Duhamel correlation function by
$$
(f(\bm \sigma);g(\bm \sigma))_\beta^{\bm J}:= (f(\bm \sigma), g(\bm \sigma))_\beta ^{\bm J}- \langle f(\bm \sigma)\rangle_\beta ^{\bm J}\langle g(\bm \sigma) \rangle_\beta ^{\bm J}.$$
Note that the derivative of the expectation value is represented in terms of the truncated Duhamel function
$$
\frac{\partial}{\partial \mu_1^w} \langle f(\bm \sigma) \rangle_\beta^{\bm J}= \beta |\Lambda_L| (f(\bm \sigma);o^{w})_\beta^{\bm J}.
$$

\section{Nishimori's gauge theory}
Nishimori's gauge theory can be extended to disordered quantum XYZ mixed $p$-spin glass models \cite{IS}. 
Let us define a gauge transformation in Nishimori's gauge theory for quantum spin glass models \cite{N,MNC,CGN}.
For a certain fixed $u (=x,y,z)$, define a unitary operator
$U_u(\tau):=\prod_{j\in \Lambda_L}(\sigma_j^u)^{(1-\tau_j)/2}$
for a spin configuration $\tau: \Lambda_L \to \{1,-1\}$.  
Define a gauge transformation with
$U_u(\tau)$ of spin operators and couplings for $w\neq u$ by
\begin{equation}
\sigma_i^w \to  \tau_i\sigma_i^w=U_u(\tau)\sigma_i^wU_u(\tau)^\dag,
\ \ \ \ J_{X,p}^{w} \to  J_{X,p}^{w} \tau_X.
\end{equation}
The Hamiltonian is invariant under the gauge transformation with any $U_w(\tau)$.
For example, for $u=x$ and $w=y,z$, the Hamiltonian is invariant  in the following
\begin{equation}
H(\bm \sigma^x,\tau\bm \sigma^y,\tau\bm \sigma^z,\bm J ^x,\bm J^y \tau, \bm J^z \tau)
= H( \bm \sigma^x, \bm \sigma^y,\bm \sigma^z,\bm J^x,\bm J^y ,\bm J^z ).
\label{gaugeinv}
\end{equation}
The distribution function is transformed in the following covariant form
\begin{equation}
P_p^{w} (J_{X,p}^{w} \tau_X)= P_p^{w} (J_{X,p}^{w}) e^{\frac{\mu_p^{w}}{{\Delta_p^{w}}^2} J_{X,p}^{w} (\tau_X-1) },
\label{gaugecov}
\end{equation}
Let $u$ be one of $x,y,z$ and  
 $v, w $ be other two of them.
 Define 
Nishimori's inverse temperature $\beta_{p}^u$ with respect to the gauge transformation $U_u(\tau)$ 
by 
\begin{equation}
\beta_{p}^u
:=\sqrt{ \frac{{\mu_p^{v}}^2}{{\Delta_p^{v}}^2}+ \frac{{\mu_p^{w}}^2}{{\Delta_p^{w}}^2}} 
. \label{betaNx}
\end{equation}
 Define  i.i.d. Gaussian r.v.s  $(K_{X,p}^u)_{X\in {\cal B}_p}$ and $(G_{X,p}^u)_{X\in {\cal B}_p} $ by
\begin{equation}
K_{X,p}^u:= \frac{1}{
\beta_{p}^u
} \Big( \frac{\mu_p^{v}}{{\Delta_p^{v}}^2} J_{X,p}^{v}+ \frac{\mu_p^{w}}{{\Delta_p^{w}}^2}J_{X,p}^{w}\Big), \ \ \ 
G_{X,p}^u:= \frac{
\mu_p^{w} J_{X,p}^{v}- \mu_p^{v}J_{X,p}^{w}}{\beta_p^u\Delta_p^v\Delta_p^w},
\label{KK'}
\end{equation}
which satisfy $${\mathbb E} K_{X,p}^u = 
\beta_{p}^u
, \ \  {\mathbb E} (K_{X,p}^u- 
\beta_{p}^u
)^2 =1, \ \ \ \  
{\mathbb E} G_{X,p}^u = 0, \ \  {\mathbb E} {G_{X,p}^u}^2 =1, \ \ \ \  
{\mathbb E}K_{X,p}^u G_{X,p}^u = 0. 
$$ 
Note 
the following relation of random variables
$$
(K_{X,p}^u-
\beta_{p}^u
)^2+{G_{X,p}^u}^2 = \frac{1}{{\Delta_p^v}^2}(J_{X,p}^v -\mu_p^v)^2+\frac{1}{{\Delta_p^w}^2}(J_{X,p}^w -\mu_p^w)^2.
$$
Therefore, the distribution function of $J_{X,p}^v, J_{X,p}^w$ is represented in terms of $K_{X,p}^u, G_{X,p}^u$
$$
 P_p^v(J_{X,p}^v) P_p^w(J_{X,p}^w) dJ_{X,p}^vdJ_{X,p}^w= \frac{1}{2 \pi}\exp \Big[
-\frac{1}{2}(K_{X,p}^u- 
\beta_{p}^u
)^2-\frac{1}{2}{G_{X,p}^u}^2 \Big]dK_{X,p}^udG_{X,p}^u.
$$
A sequence of Nishimori's inverse temperatures for all $p\in {\cal P}$ is denoted by
\begin{equation}
{\bm \beta}_{\rm N}^u:= (\beta_p^u)_{p\in {\cal P}}.\label{betaN}
\end{equation}
Define the following classical Hamiltonian for a spin configuration $\tau: \Lambda_L \to \{1,-1\}$
\begin{equation}
H_{\rm cl}(\tau, \bm K^u):=-\sum_{p\in {\cal P}} \sum_{X\in {\cal B}_p}K_{X,p}^u\tau_X.
\label{Hamilc}
\end{equation}
Several equalities among correlation functions \cite{MON} are extended in the following.   
{\lemma \label{NMq} 
Let $u$ be one of $x,y,z$ and $w (\neq u)$ be another one. 
 The one point function for any $X \subset \Lambda_L$ satisfies
\begin{equation}
\mathbb E \langle \sigma_X^w \rangle_\beta^{\bm J}=  \mathbb E \langle \sigma_X^w  \rangle_\beta^{\bm J} \langle \tau_X  \rangle_{{\bm \beta}_{\rm N}^u}^{(\bm 0,\bm K^u)},
\label{1pointNMq}
\end{equation}
and two point  functions for any $X,Y \subset \Lambda_L$  satisfy 
\begin{equation}
\mathbb  E \langle \sigma_X^w \rangle_\beta^{\bm J} \langle  \sigma_Y^w \rangle_\beta^{\bm J} =  \mathbb E \langle \sigma_X^w \rangle_\beta^{\bm J} \langle 
\sigma_Y^w \rangle_\beta^{\bm J} \langle \tau_X \tau_Y  \rangle_{{\bm \beta}_{\rm N}^u}^{(\bm 0,\bm K^u)},
\ \ \ \mathbb E \langle \sigma_X^w \sigma_Y^w \rangle_\beta^{\bm J} =  \mathbb E
 \langle \sigma_X^w \sigma_Y^w   \rangle_\beta^{\bm J} \langle \tau_X \tau_Y   \rangle_{{\bm \beta}_{\rm N}^u}^{(\bm 0,\bm K^u)}. \label{2pointNMq}
\end{equation}
Also Duhamel function and truncated Duhamel function
satisfy
\begin{equation}
 \mathbb E (\sigma_X^w, \sigma_Y ^w)_\beta^{\bm J} =  \mathbb E
 ( \sigma_X^w, \sigma_Y^w   )_\beta^{\bm J}  \langle \tau_X \tau_Y   \rangle_{{\bm \beta}_{\rm N}^u}^{(\bm 0,\bm K^u)}, \ \ \ \ 
 \mathbb E (\sigma_X^w; \sigma_Y ^w)_\beta^{\bm J} =  \mathbb E
 ( \sigma_X^w; \sigma_Y^w   )_\beta^{\bm J}  \langle \tau_X \tau_Y   \rangle_{{\bm \beta}_{\rm N}^u}^{(\bm 0,\bm K^u)}. \label{DuhamelNM}
\end{equation}
Multiple point functions satisfy corresponding  extended formulae.\\
\\
Proof. } Here, we prove these relations for a specific case $u=x$ and $w=z$ or $w=y$ for simplicity. Extensions to other components are straightforward.   
For $X \subset \Lambda_L$ 
the one point function $\mathbb E \langle \sigma_X^z\rangle_\beta^{\bm J}$ written in the integration over
$\bm J$ can be represented in terms of gauge transformed form using the gauge invariance (\ref{gaugeinv}) of the Hamiltonian  
and the gauge covariance (\ref{gaugecov}) of the distribution.
\begin{eqnarray}
&&\mathbb E \langle \sigma_X^z \rangle _\beta^{\bm J} = 
\int \langle \sigma_X^z  \rangle_\beta^{\bm J} 
\prod_{p\in {\cal P}} \prod_{Y \in {\cal B}_p}\prod_{w=x,y,z}P_p^{w}(J_{Y,p}^{w})dJ_{Y,p}^{w} 
 \nonumber \\
  &&=
  \int  \langle \sigma_X^z  \rangle_\beta^{\bm J} \tau_X \prod_{p\in {\cal P}}
\prod_{Y \in {\cal B}_p}P_p^{x}(J^{x}_{Y,p})  dJ_{Y,p}^{x}\prod_{w=y,z}P_p^{w}(J_{Y,p}^{w} \tau_Y)dJ_{Y,p}^{w}
   \nonumber \\
  &&= \int \langle \sigma_X^z  \rangle_\beta^{\bm J} \tau_X  \prod_{p\in {\cal P}} \prod_{Y\in {\cal B}_p}
  e^{(\frac{\mu_p^{y}}{{\Delta_p^{y}}^2}J_{Y,p}^{y}+\frac{\mu_p^{z}}{{\Delta_p^{z}}^2}J_{Y,p}^{z})(\tau_Y-1)}\prod_{w=x,y,z}P_p^{w}(J_{Y,p}^{w})
  dJ_Y^{w}. \nonumber 
  \end{eqnarray} 
Then, rewrite the one point function in terms of $\bm K_p^x$ and  ${\bm \beta}_{\rm N}^x$ defined by (\ref{KK'}), (\ref{betaN})
 \begin{eqnarray}
&&\mathbb E \langle \sigma_X^z \rangle _\beta^{\bm J} \nonumber \\
&&= 2^{-|\Lambda_L|}\sum_{\tau \in \{1,-1\}^{\Lambda_L}} \int \langle \sigma_X^z  \rangle_\beta^{\bm J} \tau_X
   e^{\sum_{p\in {\cal P}}\sum_{Y \in {\cal B}_p}\beta_p^x
 K_{Y,p}^x(\tau_Y-1) }
\prod_{p\in {\cal P}} \prod_{Y \in {\cal B}_p} \prod_{w=x,y,z} P_p^{w} (J_{Y,p}^{w}) dJ_{Y,p}^{w}  \nonumber \\
 &&= 2^{-|\Lambda_L|}\int \langle \sigma_X^z  \rangle_\beta^{\bm J}  \langle  \tau_X  \rangle_{{\bm \beta}_{\rm N}^x}^{(\bm 0,\bm K^x)}
\sum_{\xi \in \{1,-1\}^{\Lambda_L}}  e^{\sum_{p\in {\cal P}}\sum_{Y\in {\cal B}_p}\beta_p^x
 K_{Y,p}^x(\xi_Y-1) } 
\prod_{p,Y,w} P_{w} (J_{Y,p}^{w})    dJ_{Y,p}^{w} \nonumber \\
  &&= 2^{-|\Lambda_L|}\sum_{\xi \in \{1,-1\}^{\Lambda_L}} \int \langle \sigma_X^z  \rangle_\beta^{\bm J}  \langle  \tau_X  \rangle_{{\bm \beta}_{\rm N}^x}^{(\bm 0,\bm K^x)}
\prod_{p\in {\cal P}}\prod_{Y \in {\cal B}_p} P_p^{x} (J_{Y,p}^{x} )  dJ_{Y,p}^{x}\prod_{w=y,z}  P_p^{w} (J_{Y,p}^{w}\xi_Y )  dJ_{Y,p}^{w}
 \nonumber \\
  &&= 2^{-|\Lambda_L|}\sum_{\xi \in \{1,-1\}^{\Lambda_L}} \int \langle \sigma_X^z  \rangle_\beta^{\bm J}  \langle  \tau_X  \rangle_{{\bm \beta}_{\rm N}^x}^{(\bm 0,\bm K^x)}
\prod_{p\in {\cal P}} \prod_{Y \in {\cal B}_p}\prod_{w=x,y,z}  P_p^{w} (J_{Y,p}^{w})  dJ_{Y,p}^{w} \nonumber \\
&&=\mathbb E \langle \sigma_X^z  \rangle_\beta^{\bm J}  \langle  \tau_X  \rangle_{{\bm \beta}_{\rm N}^x}^{(\bm 0,\bm K^x)},
\label{1point-gauge}
\end{eqnarray}
where $\{1,-1\}^{\Lambda_L}$ denotes  a set of all spin configurations $\tau: \Lambda_L \to \{1,-1\}$, and 
we have used the gauge
 invariance of 
$\langle \sigma_X^z  \rangle_\beta^{\bm J}  \langle  \tau_X  \rangle_{{\bm \beta}_{\rm N}^x}^{(\bm 0,\bm K^x)}$ under 
another gauge transformation 
$\sigma_X^w \to \xi_X\sigma_X^w, \  \tau_X \to \xi_X\tau_X ,\  J^w_{X,p} \to J^{w}_{X,p}\xi_X $ for $w=y,z$ 
 to obtain the last line. The equality for $u=y$
 $$
\mathbb E \langle \sigma_X^y  \rangle_\beta^{\bm J}=\mathbb E \langle \sigma_X^y  \rangle_\beta^{\bm J}  \langle  \tau_X  \rangle_{{\bm \beta}_{\rm N}^x}^{(\bm 0,\bm K^x)},
 $$
  is obtained in the same procedure.
Other identities of  multiple point functions are obtained in the same way. $\Box$  

\section{$\mathbb Z_2$-symmetry breaking in the spin glass transition}
Here, we study spin glass phase transition in the quantum XYZ mixed even $p$-spin glass model
 with coupling constants $(\beta, \bm \Delta,\bm \mu)$. 
For a certain fixed $u (=x,y,z)$,
 the Hamiltonian (\ref{Hamilq}) is invariant under the unitary transformation 
 $U_u:=\sigma_{\Lambda_L}^u$, if $\Delta_p^w=\mu_p^w=0$ for  all odd $p \in {\cal P}$ and for all $w (\neq u)$.  
 Consider the  Hamiltonian 
which is invariant under 
all $U_x,U_y,U_z$, and apply
  a symmetry breaking field $\bm J_1^w$ for a certain fixed $w \ (\neq u)$ 
  with coupling constants $(\Delta_1^w, \mu_1^w)$ to this invariant Hamiltonian.  
Let ${\cal P}_2$ be a set  of positive even integers, and define a set ${\cal P}$ by
 \begin{equation}
 {\cal P}:=\{1\} \cup {\cal P}_2,\label{evenp}
 \end{equation} 
 to define the Hamiltonian (\ref{Hamilq}). Note the identity ${\cal B}_1 = \Lambda_L$.
 Define a function of $\bm \mu_1$ and $L$ as a finite size ferromagnetic order parameter
\begin{equation}
m_L^w(\bm \mu_1) :=\frac{1}{\beta} \frac{\partial}{\partial \mu_1^w} p_L(\beta, \bm \Delta, \bm \mu)
=\mathbb E \langle o^{w} \rangle_\beta^{\bm J}
=\frac{1}{|\Lambda_L|} \sum_{i\in \Lambda_L} \mathbb E\langle  \sigma_i^w \rangle_\beta ^{\bm J}.
\end{equation}
 Let $\bm e^w$ be a unit vector in the $w$ direction, and consider the $\mathbb Z_2$-symmetry breaking  field  
 with $(\bm \Delta_1, \bm \mu_1) = (\Delta_1^w \bm e^w, \mu_1^w \bm e^w)$.
 The following $\mathbb Z_2$-symmetric limit 
 $$ m^w := \lim_{(\Delta_1^w,\mu_1^w) \to (0,0)}  \lim_{L \to \infty} m_L^w(\bm \mu_1),
 $$
 defines the spontaneous magnetization, which  measures the spontaneous $\mathbb Z_2$-symmetry breaking.
To define three magnetic phases, define spin glass order parameter by
 $$
q^w:= \lim_{(\Delta_1^w,  \mu_1^w) \to (0,0)} \lim_{L \to \infty} \frac{1}{|\Lambda_L|}\sum_{i\in \Lambda_L}\mathbb E (\langle \sigma_i^w \rangle_\beta^{\bm J})^2.
$$
Note that $(m^w) ^2 \leq q^w$ for any $w$.  Since 
the $\mathbb Z_2$-symmetry defined by $U_u=\sigma_{\Lambda_L}^u \ (u\neq w)$ implies $\langle \sigma_i^w\rangle_\beta^{\bm J}=0$, 
also the order parameter $q^w$ measures the spontaneous $\mathbb Z_2$-symmetry breaking.
Define vector order parameters by $\bm  m:=(m^x,m^y,m^z)$, $\bm  q :=(q^x,q^y,q^z)$. 
The paramagnetic phase is defined by $\bm  q= \bm  0 = \bm  m$.
The spin glass phase is defined by $\bm  m=\bm  0$, $\bm  q \neq \bm  0$.
The ferromagnetic phase is defined by $\bm  m \neq \bm  0, \bm  q \neq \bm  0$.
All three $\mathbb Z_2$-symmetries are preserved in the paramagnetic phase. At least one of 
them is broken in the spin glass and the ferromagnetic phases. 

\paragraph{Assumptions }{\it  Consider the disordered quantum XYZ mixed even $p$-spin model 
with three $\mathbb Z_2$-symmetries defined by the unitary operators $U_u:=\sigma_{\Lambda_L}^u \ (u=x,y,z)$,  
in a $\mathbb Z_2$-symmetry breaking field 
$\bm J_1^w$ with coupling constants $(\Delta_1^w, \mu_1^w)$ for $w \ (= x,y,z)$.
Assume the following  A1 on the corresponding classical model and A2 on the quantum XYZ mixed $p$-spin glass model with $\Delta_1^w =0$.
\\
\\
A1. 
In the paramagnetic phase of the classical model defined by (\ref{Hamilc}),  the following function is bounded 
\begin{equation} 
 \frac{1}{|\Lambda_L|} \sum_{i,j\in \Lambda_L} \sqrt{ \mathbb  E \langle \tau_i \tau_j  \rangle_{{\bm \beta}_{\rm N}^u}^{(\bm 0,\bm K^u)}}
\leq C,
\end{equation} 
 by  a positive number $C$ depending on coupling constants and independent of $L$.
 \\
 \\
A2.  For $\bm \Delta_1=\bm 0$, 
the finite size ferromagnetic susceptibility  is bounded by 
\begin{equation}
\Big|\frac{\partial {m_L^w}}{\partial \mu_1^v}( \bm \mu_1 ) \Big|\leq \Big|\frac{\partial  {m_L^w}}{\partial \mu_1^v}(\bm 0)\Big| \label{GHS}
\end{equation}
for any $v,w (=x,y,z)$ for sufficiently small $|\bm \mu_1|$ for sufficiently large $L$. 
}\\
\\
A1 is valid, if the function $\mathbb E \langle \tau_i \tau_j \rangle_{{\bm \beta}_{\rm N}^u}^{(\bm 0,\bm K^u)}$  decays exponentially
for $|i-j| \gg 1$ 
in the paramagnetic phase of the classical  model for $\bm Q= \bm 0,  \bm J_1=\bm 0$. 
\\
A2 implies that the singular behavior of the susceptibility for $\bm \mu_1\neq \bm 0$ becomes weaker than that for $\bm \mu_1=\bm 0$.
This is equivalent to the fact that the finite size nonlinear susceptibility is non-positive 
$
\frac{\partial^3 {m_L^w}}{\partial {\mu_1^v}^3}(\bm 0) \leq 0,
$  
at $\bm \Delta_1=\bm \mu_1=\bm 0$, since $\frac{\partial^2 {m_L^w}}{\partial {\mu_1^v}^2}(\bm0)=0$ by the $\mathbb Z_2$-symmetry. 
In the nearest neighbor ferromagnetic Ising model with $\bm Q=
\bm 0,\bm J_1 = \bm 0
$, 
this non-positivity is guaranteed by the Lebowitz inequality \cite{L}.
In this model, the inequality (\ref{GHS}) is always valid for any $\mu_1^v > 0$ by the  Griffiths-Hurst-Sharman inequality \cite{GHS}.

\subsection{Absence of spontaneous magnetization}
First, absence of spontaneous ferromagnetic  magnetization is proven.

{\theorem \label{corq}  
Consider the quantum XYZ mixed even $p$-spin glass model defined by the  Hamiltonian 
(\ref{Hamilq})  with 
a symmetry breaking field $\bm J_1$ with $(\bm \Delta_1,\bm \mu_1)$. 
The spontaneous ferromagnetic magnetization vanishes
\begin{equation}
 \bm m=\bm 0,
\end{equation}
for any $\beta$ in the model with $\bm J$, if the corresponding classical model
with random exchanges $ \bm K^u$ defined by  
 (\ref{KK'}) is in the paramagnetic phase.\\
\\
Proof.} This is proven using Lemma \ref{NMq}. Consider a magnetization process in the quantum model for $\beta>0$ and 
${\bm \beta}_{\rm N}^u$ defined by (\ref{betaN}). 
The identity (\ref{1pointNMq}) gives a bound on the  magnetization. 
\begin{eqnarray*}
|{\mathbb E} \langle \sigma_i^w \rangle_{\beta}^{\bm J}|
&=&|{\mathbb E} \langle \sigma_i^w \rangle_{\beta}^{\bm J}\langle \tau_i \rangle_{{\bm \beta}_{\rm N}^u}^{(\bm 0,\bm K^u)}  |
\leq {\mathbb E}| \langle \sigma_i^w \rangle_{\beta}^{\bm J} ||\langle \tau_i \rangle_{{\bm \beta}_{\rm N}^u}^{(\bm 0,\bm K^u)}  |
\leq {\mathbb E}|\langle \tau_i \rangle_{{\bm \beta}_{\rm N}^u }^{(\bm 0,\bm K^u)} |\\
&\leq& \sqrt{{\mathbb E}( \langle \tau_i \rangle_{{\bm \beta}_{\rm N}^u}^{(\bm 0,\bm K^u)} )^2}
 =\sqrt{{\mathbb E} \langle \tau_i \rangle_{{\bm \beta}_{\rm N}^u }^{(\bm 0,\bm K^u)}},
\end{eqnarray*}
for any $i \in \Lambda_L$. This and  Jensen's inequality imply
\begin{equation}
\mathbb E   \langle o^w \rangle_{\beta}^{\bm J}
=\frac{1}{|\Lambda_L|} \sum_{i\in \Lambda_L} 
{\mathbb E} \langle \sigma_i ^w\rangle_{\beta}^{\bm J} \leq 
 \frac{1}{|\Lambda_L|} \sum_{i\in \Lambda_L} 
\sqrt{{\mathbb E} \langle \tau_i \rangle_{{\bm \beta}_{\rm N}^u}^{(\bm 0,\bm K^u)}} \leq 
\sqrt{ \frac{1}{|\Lambda_L|} \sum_{i\in \Lambda_L} 
{\mathbb E} \langle \tau_i \rangle_{{\bm \beta}_{\rm N}^u}^{(\bm 0,\bm K^u)}}.
\label{boundmagnetization}
\end{equation}
If the classical model with  $(\bm 0, \bm K^u)$ and ${\bm \beta}_{\rm N}^u$  
has no spontaneous magnetization for two of  $u=x,y,z,$
the right hand side of (\ref{boundmagnetization}) vanishes in the limit $\beta_1^u\to 0$. Therefore, the magnetization 
vanishes  in this limit
\begin{equation}
m^w=\lim_{(\Delta_1^w,\mu_1^w) \to (0,0)}\lim_{L\to \infty}  \mathbb E   \langle o^{w} \rangle_{\beta}^{\bm J}
 =0,
\end{equation}
also in the quantum XYZ $p$-spin glass model with the random exchanges $\bm J$ 
for any $\beta >0$  for any  $w (=x,y,z)$. This completes the proof. $\Box$
\\
\\
%

\subsection{Bound on the susceptibility}
Finally, let us explain that the ferromagnetic susceptibility has a finite upper bound  under the
acceptable assumptions A1, A2 in the $\mathbb Z_2$-symmetry breaking 
phase transition of the quantum XYZ mixed even $p$-spin glass model. 
For $\bm \Delta_1=\bm 0$, 
regard the sample expectation of the magnetization as a function of the deterministic field $\bm \mu_1$
and the system size $L$.
Define a magnetic susceptibility  in the infinite-volume limit by  
\begin{equation}
\chi^{v,w}(\bm \mu_1):=
\frac{\partial }{\partial \mu_1^v} \lim_{L\to\infty}
m_L^w(\bm \mu_1). 
\end{equation}
The following theorem states a boundedness of the susceptibility for $\bm \mu_1 = \bm 0.$

{\theorem \label{sus}  
Consider the quantum XYZ mixed even $p$-spin glass model under a symmetry breaking field $\bm J_1$ with
$\bm \Delta_1=\bm 0$ satisfying assumptions A1 and A2. If the corresponding classical model with $\bm K^u$ defined by (\ref{KK'})
at ${\bm \beta}_{\rm N}^u$ defined by (\ref{betaN}) is in paramagnetic phase,
 then the magnitude of magnetic susceptibility in the quantum XYZ spin glass model
  at $\bm \mu_1=\bm 0$
 is bounded from the above 
 \begin{equation}
|\chi^{v,w}(\bm 0)| \leq 2\beta C.
\end{equation} 
for any temperature for any $v,w=x,y,z$.
\\

\noindent
Proof.}   
Since the function $p(\beta, \bm \Delta, \bm \mu)$ exists as a convex function of $\mu_1^w$ for any $w$,  
$p(\beta, \bm \Delta, \bm \mu)$ is continuously differentiable at almost all  $\mu_1^w$,
 where the infinite volume limit of the ferromagnetic magnetization 
\begin{equation}
\lim_{L\to\infty}m_L^w(\bm \mu_1)
= \frac{1}{\beta}\frac{\partial }{\partial \mu_1^w} p(\beta, \bm \Delta, \bm \mu),
\end{equation}
 is represented in terms of the partial derivative of $p(\beta, \bm \Delta, \bm \mu)$ with respect to $\mu_1^w$  \cite{I2}.
 Consider a single argument function 
 $
 m_L^w(
 t \bm e^v)$ of $t \in (0,1)$,
 where $\bm e^v$ is a unit vector in the $v$ direction. 
The mean value theorem implies that  
there exists a positive number $ \theta < 1$ for $t$,  such that  
\begin{equation}
\frac{1}{t} [m_L^w(t \bm e^v)-m_L^w(\bm 0)] =\frac{d}{dt} {m_L^w}(\theta t \bm e^v)=\frac{\partial m_L^w}{\partial \mu_1^v} (\theta  t\bm e^v).
\end{equation}
Note that  the $\mathbb Z_2$-symmetry guarantees $m_L^w(\bm 0)=0$ for any $L$. 
The assumption A2 implies that the right hand side in the above identity is bounded by
\begin{equation}
\Big| \frac{\partial m_L^w}{\partial \mu_1^v} (\theta t \bm e^v)\Big|
 \leq\Big|\frac{\partial m_L^w}{\partial \mu_1^v} (\bm 0)\Big|,
\end{equation}
 for a sufficiently small $t>0$
and for sufficiently large $L$.
For $\bm \mu_1=\bm 0$,  
the following bound on the susceptibility is obtained  
using the  identity
 (\ref{DuhamelNM}), Jensen's inequality and identity (\ref{2pointNMq}) for ${\bm \beta}_{\rm N}^u$ for any $v \neq u\neq u$ 
\begin{eqnarray}
&&
\Big|\frac{\partial m_L^w}{\partial \mu_1^v} (\bm 0)\Big|=
\frac{\beta}{|\Lambda_L|}\Big|\sum_{i,j\in \Lambda_L}\mathbb  E 
( \sigma_i^w; \sigma_j^v )_\beta^{\bm J}\Big|
 =\frac{\beta}{|\Lambda_L|}\Big|\sum_{i, j\in \Lambda_L}  \mathbb  E 
 ( \sigma_i ^w; \sigma_j^v )_\beta^{\bm J} 
 \langle \tau_i \tau_j  \rangle_{{\bm \beta}_{\rm N}^u}^{(\bm 0,\bm K^u)} \Big|
 \nonumber \\
&&\leq \frac{\beta}{|\Lambda_L|} \sum_{i,j\in \Lambda_L}  \mathbb  E| ( \sigma_i^w; \sigma_j ^v)_\beta^{\bm J}
 || \langle \tau_i \tau_j  \rangle_{{\bm \beta}_{\rm N}^u}^{(\bm 0,\bm K^u)}| 
 \leq \frac{2\beta}{|\Lambda_L|} \sum_{i,j\in \Lambda_L}  \mathbb  E| \langle \tau_i \tau_j  \rangle_{{\bm \beta}_{\rm N}^u}^{(\bm 0,\bm K^u)}| \nonumber \\
 &&\leq\frac{2\beta}{|\Lambda_L|} \sum_{i,j\in \Lambda_L}\sqrt{  \mathbb  E (\langle \tau_i \tau_j  \rangle_{{\bm \beta}_{\rm N}^u}^{(\bm 0,\bm K^u)})^2}
  = \frac{2\beta}{|\Lambda_L|} \sum_{i,j\in \Lambda_L}\sqrt{  \mathbb  E \langle \tau_i \tau_j  \rangle_{{\bm \beta}_{\rm N}^u}^{(\bm 0,\bm K^u)}}.
 \label{susq}
\end{eqnarray}
The assumption A1 on the correlation function at ${\bm \beta}_{\rm N}^u, \bm \mu_1=\bm 0$ implies 
\begin{eqnarray}
\Big|\frac{\partial m_L^w}{\partial \mu_1^v} (\bm 0)\Big|
  \leq \frac{2\beta}{|\Lambda_L|} \sum_{i,j\in \Lambda_L}\sqrt{  \mathbb  E \langle \tau_i \tau_j  \rangle_{{\bm \beta}_{\rm N}^u}^{(\bm 0,\bm K^u)}}
  \leq 2 \beta C,
\end{eqnarray}
where $C$ does not depend on $L$. 
Therefore, 
\begin{eqnarray}
\lim_{L\to\infty}\frac{1}{t}|m_L^w(t\bm e^v) -m_L^w(\bm 0) | \leq
\limsup_{L\to\infty}
 \frac{2\beta}{|\Lambda_L|} \sum_{i,j\in \Lambda_L}\sqrt{  \mathbb  E \langle \tau_i \tau_j  \rangle_{{\bm \beta}_{\rm N}^u}^{(\bm 0,\bm K^u)}}
  \leq 2 \beta C,
\end{eqnarray}
for sufficiently small $t > 0$.
The magnitude of susceptibility at any  $\beta>0$
 is bounded from the above
\begin{eqnarray}
&&\Big|\frac{\partial }{\partial \mu_1^v}\lim_{L\to\infty}m_L^w(\bm \mu_1)\Big|_{\bm \mu_1=\bm 0}
= \lim_{t \to 0} \lim_{L\to\infty} \frac{1}{t}|m_L^w(t \bm e^v) -m_L^w(\bm 0) |
 \leq 2\beta C.
\end{eqnarray}    
This completes the proof.   
 $\Box$
 

 \section{Summary and discussions}
In the present paper, Nishimori's gauge theory is extended to the quantum  XYZ mixed $p$-spin glass model.
The gauge transformation for spin operators is 
generated by the unitary operator $U_u:=\prod_{i\in \Lambda_L}(\sigma_i^u)^{(1-\tau_i)/2}$  for $u=x,y,z$
and $J_{X,p}^w \to J_{X,p}^w\tau_X$
with a classical Ising spin configuration $\tau: \Lambda_L \to \{1,-1\}$.
The covariance of spin operators and functions of couplings $\bm J$, and invariance of the Hamiltonian
provide Lemma \ref{NMq}, which claims that the sample expectation of any correlation function in the quantum XYZ spin glass model
is identical to the expectation of the original correlation function times the correlation function in the corresponding classical model on the Nishimori
 line. 
  We have discussed spontaneous $\mathbb Z_2$-symmetry breaking phenomena in the quantum XYZ mixed even $p$-spin glass model. 
 The identity  for the one point function in Lemma \ref{NMq}  enables us to prove Theorem \ref{corq}.
This gives a sufficient condition 
for that all components of the spontaneous magnetization vanish
$$
 m^w=\lim_{(\Delta_1^w,\mu_1^w) \to (0,0)} \lim_{L\to \infty}\frac{1}{|\Lambda_L| }\sum_{i\in \Lambda_L}\mathbb E \langle  \sigma_i^w \rangle_{\beta}^{\bm J} = 0,
$$ 
 in the quantum  XYZ mixed even $p$-spin glass model.
If the corresponding classical models at ${\bm \beta}_{\rm N}^u$ for each $u \neq w$ is in the paramagnetic phase, then
the quantum XYZ mixed even $p$-spin glass model at any temperature is in the paramagnetic or spin glass phase, where all components of 
the spontaneous magnetization vanish $m^x=m^y=m^z=0$. 
The spin glass transition occurs from the paramagnetic phase to the spin glass phase  by lowering temperature. 
The identity for the truncated Duhamel two point function in Lemma \ref{NMq} and assumptions A1, A2 enable us to prove  Theorem \ref{sus} for
the quantum XYZ mixed even $p$-spin glass model. Theorem  \ref{sus} claims the well-known fact that the magnetic susceptibility does not diverge in the spin glass transition between the paramagnetic and spin glass phases in a certain region of coupling constants, where the corresponding classical
mixed $p$-spin glass model is in the paramagnetic phase.

\begin{figure}[H]
\begin{center}
\includegraphics[width=50mm, angle=90]{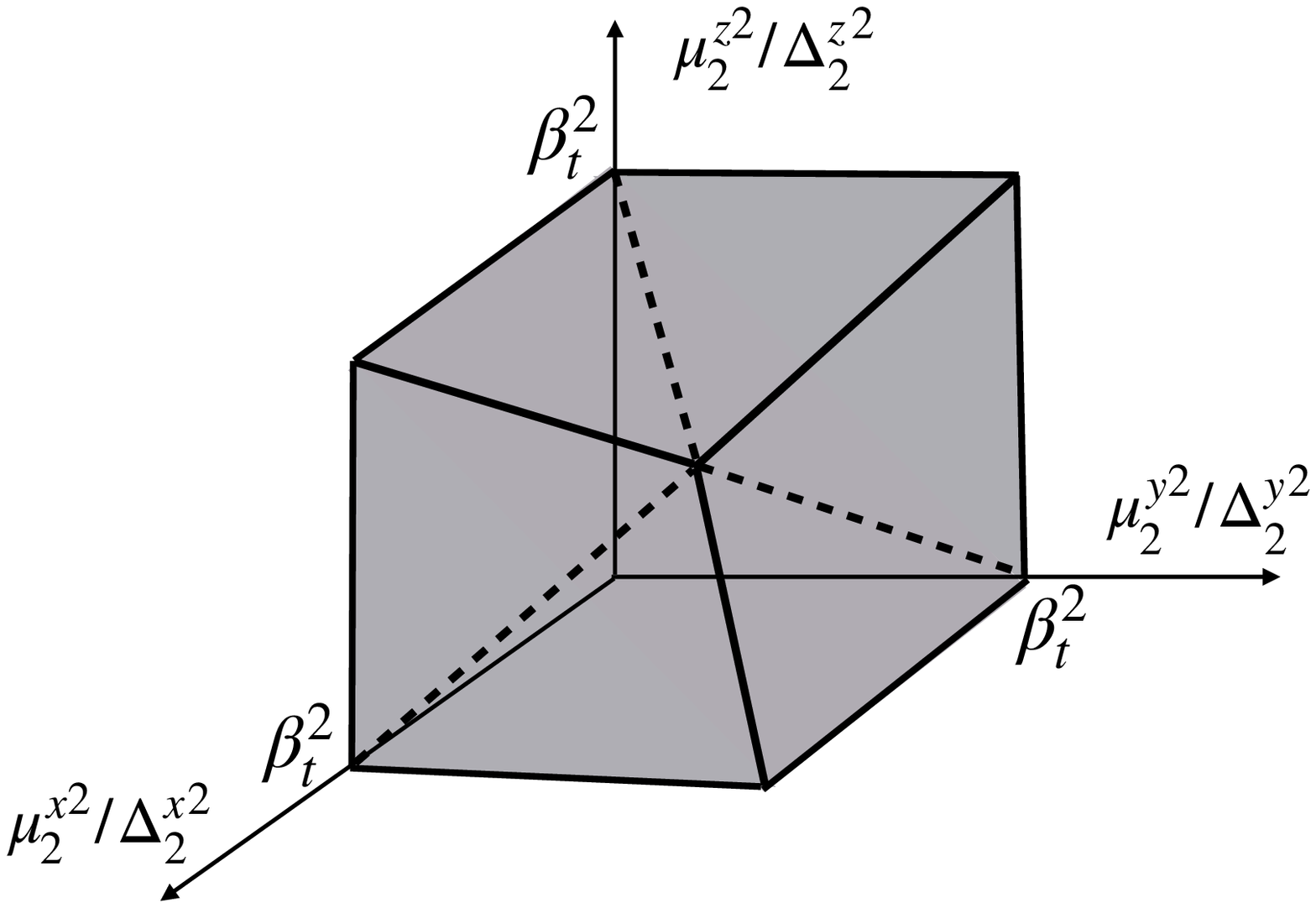}
\end{center}
\caption{The gray region depicts $S^x \cup S^y \cup S^z$, 
which is predicted to be the spin glass region
of the quantum Heisenberg XYZ spin glass model
in sufficiently low temperature. All $S^u \ (u=x,y,z)$ are congruent square-based pyramids in this coordinate system 
of the coupling constant space, then  
the boundary of $S^x \cup S^y \cup S^z$ consists of triangles and squares.  
Solid and dashed straight lines represent convex and concave edges, respectively. 
}
\label{figXYZ}
\end{figure}

Here, we discuss properties of spin glass transitions for a specific case of $p=2$ with nearest neighbor exchange interactions, which is the
quantum Heisenberg XYZ spin glass model. In this model, the collection of nearest neighbor bonds  is defined by
$$
{\cal B}_2 := \{ \{i,j\}| i,j\in \Lambda_L, |i-j|=1 \}.
$$
Its corresponding classical model is the well-known
 Edwards-Anderson (EA) model, whose phase diagram is considered to be established. 
 The EA model has the unique triple point $(\beta_t, \mu_t)$, where the paramagnetic, spin glass and ferromagnetic phases coexist.  
Define
a subspace $S^u$  $(u=x,y,z)$ of  the coupling constant space by
\begin{equation}
S^u:=\{(\Delta_2^x,\Delta_2^y,\Delta_2^z, \mu_2^x,  \mu_2^y,  \mu_2^z) | \beta_2^v, \beta_2^w < \beta_t, w \neq u \neq v \neq w \},\label{subspacedef}
\end{equation}
where the definition (\ref{betaNx}) 
for $p=2$ gives the following explicit form of the condition
\begin{equation}
 \frac{{\mu_2^{w}}^2}{{\Delta_2^{w}}^2}+\frac{{\mu_2^{u}}^2}{{\Delta_2^{u}}^2}
 =:{\beta_2^v}^2< \beta_t^2, \ \ \ 
 \frac{{\mu_2^{u}}^2}{{\Delta_2^{u}}^2}+ \frac{{\mu_2^{v}}^2}{{\Delta_2^{v}}^2}=:{\beta_2^w}^2 < \beta_t^2, \ \ \ 
\end{equation}
for $w \neq u \neq v \neq w$. Theorem \ref{corq} leads that 
if  $\beta_2^w, \beta_2^v <\beta_t$ are satisfied for any two $v,w$ of $x,y,z$, 
then all components of spontaneous magnetization vanish
 \begin{equation}m^x=m^y=m^z=0,\end{equation}
  at any temperature.
This fact implies that the union of the paramagnetic region $S_{\rm PM}$ 
and the spin glass region $S_{\rm SG}$ in the coupling constant space of the quantum Heisenberg XYZ
spin glass model at an arbitrary fixed temperature includes 
\begin{equation}
S_{\rm PM} \cup S_{\rm SG} \supset S^x \cup S^y \cup S^z, 
\label{subspace}
\end{equation}
where the right hand side is depicted in Figure \ref{figXYZ}.
For $\beta_2^u > \beta_t$, the spontaneous magnetization  can be finite $m^w\neq 0$ for $w\neq u$. 
Although $\beta_2^u <\beta_t$ and $ \beta_2^v<\beta_t$ ($u\neq v$) are sufficient condition for no spontaneous magnetization $\bm m =\bm 0$, 
there is the following natural conjecture
\begin{equation}S_{\rm SG} \subset S^x \cup S^y \cup S^z. \label{conjecture}
\end{equation}
 Namely, the region where the gauge theory proves that there is no spontaneous magnetization $\bm m=\bm 0$
 is expected to contain the spin glass region. As far as the EA model is concerned, this conjecture 
 is believed to lead to its established phase diagram. Since paramagnetic region disappears in sufficiently low temperature, this conjecture 
 leads to $S_{\rm SG} = S^x \cup S^y \cup S^z$. 
This conjecture (\ref{conjecture}) and Theorem \ref{sus} imply that
 the magnetic susceptibility is always finite in the spin glass transition of the quantum Heisenberg XYZ spin glass model. In other words, 
 this conjecture is sufficient to explain well-known experimental and numerical evidences showing the finiteness of the magnetic susceptibility in spin glass transitions.
On the other hand, what the violation of  the conjecture (\ref{conjecture}) yields is interesting to study, since  
$$S_{\rm SG} \cap (S^x \cup S^y \cup S^z)^c \neq \phi,$$
cannot be excluded rigorously. 
For example, possibility of divergent susceptibility in spin glass transition
of quantum systems should be searched. If there exists a region without spontaneous magnetization that 
cannot be proven by the gauge theory, the finiteness of the magnetic susceptibility cannot be proven either.   
The validity of the conjecture  (\ref{conjecture}) would be difficult to judge at this stage, 
but this might provide interesting problems for further studies.
\\

\noindent
{\bf Acknowledgments} \\
 C.I. is supported by JSPS (21K03393).
 \\
There is no conflict of interest. All data are provided in full in this paper.

\end{document}